\documentclass[11pt]{article}
\usepackage[margin=1in]{geometry}
\usepackage{setspace}
\usepackage{dsfont, amssymb,amsmath,amscd,latexsym, amsthm, amsxtra,amsfonts, extarrows}
\usepackage{amssymb}
\usepackage{amsmath}
\numberwithin{equation}{section}
\numberwithin{figure}{section}
\usepackage{indentfirst}
\usepackage{graphicx}
\usepackage[colorlinks=true, allcolors=blue]{hyperref}
\usepackage[utf8]{inputenc} 
\usepackage[T1]{fontenc}    
\usepackage{url}            
\usepackage{booktabs}       
\usepackage{amsfonts}       
\usepackage{nicefrac}       
\usepackage{microtype}      
\usepackage{fancyhdr}       
\usepackage{natbib}
\graphicspath{{media/}}     

\newtheorem{Definition}{Definition}
\newtheorem{Theorem}{Theorem}

\newtheorem{Proposition}{Proposition}

\usepackage{appendix}
\usepackage{subcaption}

\begin{document}

\title{\textbf{Dynamic Investment-Driven Insurance Pricing and Optimal Regulation}} 
\author{Bingzheng Chen\thanks{\ School of Economics and Management, Tsinghua University; Email: chenbzh@sem.tsinghua.edu.cn} \and Zongxia Liang\thanks{\ Department of Mathematical Sciences, Tsinghua University; Email: liangzongxia@tsinghua.edu.cn} \and Shunzhi Pang\thanks{\ School of Economics and Management, Tsinghua University; Faculty of Economics and Business, KU Leuven; Email: psz22@mails.tsinghua.edu.cn}}
\date{\today}
\maketitle
\begin{abstract}
\noindent This paper analyzes the equilibrium of insurance market in a dynamic setting, focusing on the interaction between insurers' underwriting and investment strategies. Three possible equilibrium outcomes are identified: a positive insurance market, a zero insurance market, and market failure. Our findings reveal why insurers may rationally accept underwriting losses by setting a negative safety loading while relying on investment profits, particularly when there is a negative correlation between insurance gains and financial returns. Additionally, we explore the impact of regulatory frictions, showing that while imposing a cost on investment can enhance social welfare under certain conditions, it may not always be necessary. 
\\
\vspace{0in}\\
\noindent\textbf{Keywords:} Equilibrium insurance pricing; Insurer investment; Social welfare; Insurance regulation \\ 
\bigskip
\end{abstract}

\newpage

\section{Introduction} 

The question of how insurers price their products reasonably and how regulators oversee these companies effectively has been a long-standing area of focus for both academia and industry, particularly within the field of insurance economics and actuarial science. Theoretically, insurers would not face solvency issues under the principle of actuarially fair pricing, as insurance pricing usually consists of both an actuarially fair premium and a loading premium (also known as the safety loading), ensuring that the expected profits of insurers are always non-negative.\footnote{Based on the Law of Large Numbers, the actuarially fair premium could cover expected losses, while the loading premium would account for operational costs and profit margins, allowing insurers to generate positive expected returns. } Solvency issues typically arise only in the event of unforeseen extreme risk losses or interest rate changes, but such fluctuations are inherently short-term.\footnote{This is because, after losses occur, insurers can identify the underlying causes and dynamically adjust their pricing to avoid negative expected future profits, for instance, by revising risk models or updating parameter estimations. } 

\begin{figure}[htpb]
    \centering
    \includegraphics[width=0.9\linewidth]{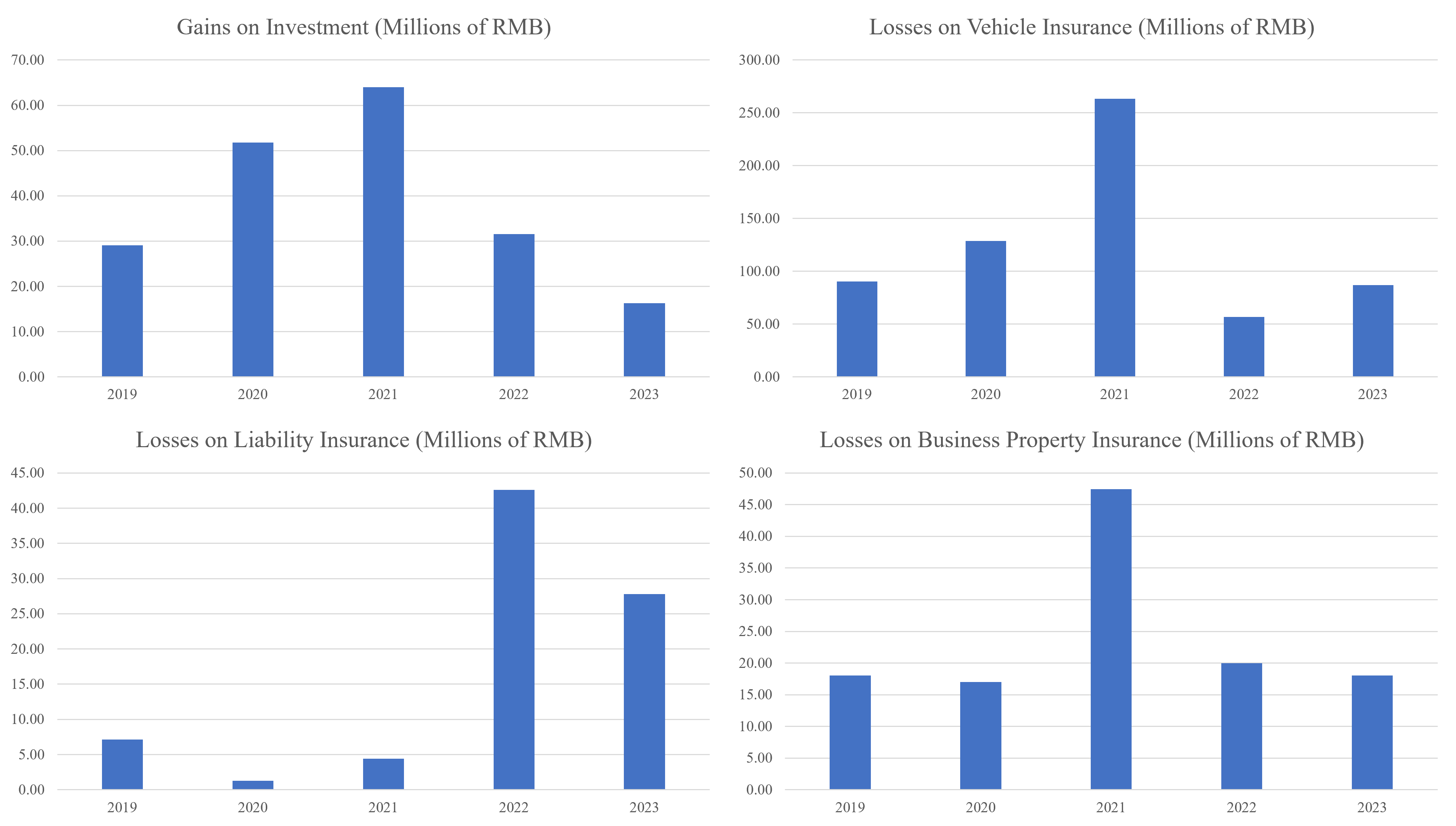}
    \caption{Investment Gains and Underwriting Losses of a Representative Insurer}
    \label{Case of Changjiang}
\end{figure}

However, recent practices reveal that underwriting losses are becoming increasingly common among insurers. For instance, in 2023, it was reported that among the 69 non-listed property and casualty (P\&C) insurers in China, 43 of them (over 60\%) experienced underwriting losses.\footnote{See https://baijiahao.baidu.com/s?id=1790310867880623315\&wfr=spider\&for=pc, where the research report was published by Thirteen Actuaries, one of China's largest insurance data consulting firms. It is reported that the median underwriting profit margin across these insurers was -3.39\%, while the simple average stood at -5.21\%, and the weighted average at -0.59\%.} Moreover, these losses appear to be persistent rather than short-term. According to a 2022 research report by the China Insurance Security Fund, over half of the P\&C insurers in China have experienced underwriting losses for five consecutive years.\footnote{The China Insurance Security Fund is a state-owned enterprise responsible for managing the insurance security fund and monitoring risk in the insurance industry. This conclusion is derived from its report China Insurance Industry Risk Assessment Report 2022.} Figure \ref{Case of Changjiang} illustrates the operating performance of a representative mid-sized insurer from 2019 to 2023.\footnote{The data is sourced from the company's annual information disclosure reports, which are publicly available and can be downloaded from https://www.cjbx.com.cn/ndxxplbg/index.jhtml. } Its three primary business lines all incurred losses every year during this period. The phenomenon of largely unprofitable underwriting operations within the P\&C industry stands in stark contrast to traditional insurance pricing theory. Understanding why and how this occurs is both important and necessary for informing industry practices. 


Despite the widespread underwriting losses, in reality, most P\&C insurers manage to generate profits through investments. For example, among the aforementioned 69 P\&C insurers in China, over 90\% reported positive investment returns in 2023.\footnote{Also see https://baijiahao.baidu.com/s?id=1790310867880623315\&wfr=spider\&for=pc, where the research report was published by Thirteen Actuaries, one of China's largest insurance data consulting firms. Statistics indicate that the median total investment return for these insurers was 3.0\%, with a simple average of 2.6\% and a maximum of 6.22\%.} Specifically, the insurer in Figure \ref{Case of Changjiang} achieved investment gains over the past five years, although the gains were insufficient to fully offset underwriting losses. Moreover, there appears to be a negative relationship between the insurer's underwriting gains and investment returns. In years with higher aggregate underwriting losses, such as 2021, the insurer recorded its highest investment returns. Conversely, in years with relatively lower aggregate underwriting losses, such as 2019 and 2023, the insurer's investment returns were also relatively lower. More precisely, it is calculated that the correlation coefficient between the insurer's investment gains and the underwriting gains of its three main business lines is -0.79 over the past five years. 

Different from banks, mutual funds, or even life insurers, the primary role of P\&C insurers should be that of risk management institutions. However, it is evident that they are increasingly behaving like other financial intermediaries, heavily reliant on investment returns. This trend is also consistent with what has been observed in the U.S. insurance market, where insurers have long been the largest institutional investors in corporate bonds \citep{koijen2023understanding}. Beyond fixed-income investments, insurers are also deeply involved in equity investments. A notable example is Warren Buffett’s investment strategy through Berkshire Hathaway, where the float generated by insurers has been used to finance large-scale equity investments. This approach has allowed Berkshire Hathaway to become one of the largest shareholders in several major corporations. 


These observations motivate us to investigate the incentives behind insurers' underwriting and investment decisions. In this paper, we adapt the dynamic framework of \citet{henriet2016dynamics} to analyze the equilibrium outcomes of the insurance market. The theoretical results derived under specific conditions indicate that insurers may rationally choose to incur underwriting losses while relying on investment profits to sustain their operations in a long-term equilibrium, driven by the diversification effect. This property provides a possible explanation for the phenomenon observed among Chinese P\&C insurers, as discussed. 

More specifically, unlike the traditional approach where the safety loading is treated as an exogenously given and positive constant, our model considers the loading as the price of insurance, endogenously determined by the interaction of market supply and demand. Such a pricing method can be referred to as the \textit{equilibrium pricing of insurance}. Within this framework, we further explore the welfare implications of the insurance market and its interplay with the financial market. Additionally, recognizing the importance of regulatory oversight, we examine the optimal regulatory policies that could be implemented under varying market conditions to address insurers' behaviors effectively. 


Several papers in the literature have explored the dynamic insurance pricing through an equilibrium approach. For instance, \citet{winter1994dynamics} argues that competitive pressures lead insurers to adjust prices based on profitability, resulting in cycles where premiums fluctuate with changing market conditions. Similarly, \citet{henriet2016dynamics} develop a continuous-time equilibrium model where insurers' aggregate net worth determines the underwriting, dividend, and recapitalization strategies. They show that equilibrium insurance prices are predictable and exhibit asymmetric reversals, explaining the industry’s underwriting cycles. Additionally,  \citet{feng2023insurance} introduce a dynamic equilibrium model where insurers differ in their beliefs about expected loss rates. The stochastic divergence in these beliefs adds volatility to insurance prices, which helps explain the cyclical behavior observed in the market. 

While these studies all conclude that insurers should realize positive expected profits in policies due to a positive safety loading in equilibrium, we propose that the equilibrium loading can naturally become negative under certain parameter settings when we introduce an external financial market for insurers to invest in and the investment returns are negatively correlated with uncertain underwriting gains.\footnote{Negative correlation between investment returns and underwriting gains is equivalent to positive correlation between investment returns and underwriting losses. In terms of expression, we also refer to this scenario as a negative correlation between the two markets or risks in this paper.} In such a scenario, insurers may be incentivized to offer negative loading premiums to attract policyholders, as the expected returns from investments outweigh the underwriting losses. By leveraging premiums as investment capital, insurers can maintain profitability despite incurring underwriting deficits.\footnote{This property was hinted at in \citet{henriet2016dynamics}, which slightly discusses the possibility of negative equilibrium loading as insurers hedge their losses through complementary financial market activities. However, they do not provide a detailed proof or specify the market conditions under which this could occur. Our work extends and refines their discussion. } 

In our setting, introducing a correlation between risks in insurance and financial markets is both natural and necessary, as financial market performance and insurance outcomes are often intertwined.\footnote{For instance, during economic downturns insurers typically experience lower investment returns coupled with higher underwriting losses, such as increased property claims or credit defaults, as seen in the 2008 financial crisis. Conversely, during economic booms, while financial markets perform strongly, insurance losses, particularly in liability and workers' compensation, may increase due to heightened economic activity, such as construction related accidents. Climate events like hurricanes not only lead to substantial insurance claims but can also disrupt financial markets, creating volatility.} Although some studies have examined how the correlation between the two risks affects insurers' investment and reinsurance behavior \citep[see][]{cheng2024robust, wu2025alpha}, few have explored its impact on underwriting and pricing strategies. When the two risks driven by Brownian motions are uncorrelated, the insurance and financial markets function as independent entities, and the insurer’s underwriting strategy remains consistent with the scenario where no investment opportunities exist. 

However, when the two risks are correlated, diverging from the traditional view that the insurer's surplus is non-tradable, we treat insurance policies as tradable assets and incorporate them into the insurer's portfolio selection. The ratio of profit margins between insurance policies and risky assets becomes the key factor in determining the insurer's behavior. When this ratio is relatively large and exceeds the correlation coefficient, insurers maintain a balance between underwriting and investment, as both yield competitive returns. But when this ratio falls below the correlation coefficient, investment becomes more profitable, leading insurers to hold zero underwriting position and shift their capital primarily toward investment. 

For a representative insurer with constant absolute risk aversion (CARA) utility, we apply dynamic programming to derive the optimal underwriting and investment strategies under a given price process. The analytical solution demonstrates that whenever the insurer is willing to underwrite, the optimal underwriting amount is a linear function of the price, which represents the market supply curve. Assuming a linear consumer demand curve within an exogenously set price range (bounded by a regulator-defined upper and lower limit), the equilibrium insurance quantity and price are determined by the intersection of the two curves. Depending on the parameter conditions (mainly the price bounds and correlation coefficient), three outcomes are possible: (1) a \textit{positive insurance market} (with positive equilibrium underwriting), (2) a \textit{zero insurance market} (with zero equilibrium underwriting), and (3) \textit{market failure} (where no equilibrium exists). 

In a positive insurance market, social welfare in equilibrium (the sum of consumer and producer surplus) can be calculated. When the two markets are correlated, comparative static analysis shows that the investment Sharpe ratio significantly affects market equilibrium. Specifically, the insurance price decreases (increases) as the Sharpe ratio rises when the two market risks are negatively (positively) correlated. Conversely, the underwriting amount decreases (increases) when the risks are positively (negatively) correlated, while investment always increases with the Sharpe ratio. As a result, social welfare decreases (increases) in the Sharpe ratio when the two market risks are positively (negatively) correlated. 

Within this framework, we further study the impact of regulatory frictions on market equilibrium and social welfare. In practice, insurers' investment behavior is constrained by risk-based capital regulation, which limits their risk-shifting incentives \citep[see][]{koijen2022fragility}. To capture this, we introduce a quadratic form of regulatory-induced investment costs into the insurer’s capital flow process and similarly solve for the equilibrium. In a positive insurance market, regulatory costs do not affect underwriting strategy, market equilibrium, or social welfare when the two markets are uncorrelated. However, when the two markets are correlated, regulatory costs would affect these outcomes, and the optimal cost that maximizes social welfare can be derived, which is not necessarily zero. Therefore, while in certain cases appropriate regulation of insurers' investment behavior can enhance social welfare in the insurance market, in other situations regulation may be unnecessary. A case-by-case analysis is required for policy makers. 

We contribute to the ongoing literature in actuarial science on optimal investment and pricing for insurers. While previous studies have examined insurers' investment strategies under various optimization objectives \citep{liu2004optimal, yang2005optimal}, with ambiguity aversion \citep{zeng2016robust, guan2019robust}, and in connection with optimal reinsurance \citep{zeng2011optimal, yi2013robust}, our focus lies in the interaction between investment and underwriting strategies. Regarding insurance pricing, while fewer studies address how premiums should be determined in a competitive market, one approach is to model it as a game between insurers, as explored in \citet{emms2005pricing}, \citet{emms2007dynamic, emms2012equilibrium}, \citet{asimit2018insurance} and \citet{li2021dynamic}. Instead, we adopt a representative insurer model and follow the equilibrium pricing framework of insurance proposed by \citet{henriet2016dynamics} and \citet{feng2023insurance}, though we believe that the game-theoretical problem within this framework merits further research. 

We also contribute to the growing literature on the financial economics of insurance. While classical insurance market theories have primarily focused on the demand side, recent studies have emphasized the importance of supply-side dynamics, highlighting the role of insurers as financial intermediaries in capital markets \citep{koijen2023financial}.\footnote{For more related literature, see \citet{ge2022financial}, \citet{egan2022conflicting}, \citet{ellul2022insurers}, \citet{sen2023regulatory}, \citet{kubitza2021investor}, \citet{kubitza2023life}, \citet{koijen2024aggregate}, \citet{knox2024insurers}, and others.} Financial and product market frictions, combined with statutory reserve regulations, can lead to the shadow cost and result in higher prices \citep{koijen2015cost, koijen2022fragility}.
While several empirical studies exist, there is limited work on the theoretical foundations of the financial economics of insurance. Our framework not only capture the dynamic equilibrium of insurance markets and explain the negative loading phenomenon, but also allow for further analysis of social welfare and regulatory implications. 


We must acknowledge the limitations of this paper. Theoretically, the prices of financial assets follow the classical Black-Scholes market assumption and are given exogenously in our model. It would be interesting to explore how insurers influence the prices of financial assets in a general equilibrium model. Empirically, due to data limitations, we do not provide a detailed investigation into the relationship between the profitability of insurers' business and investment activities. As the main focus of this paper is to derive theoretical insights, we just cite statistics from research reports and present data from a representative insurer to support our argument, which lacks rigorous empirical validation. Addressing these limitations could be a promising direction for future studies. 

The remainder of this paper is organized as follows. Section \ref{Section Model} introduces the model setup and defines market equilibrium. Section \ref{Section Equilibrium} solves the insurer’s optimization problem and conducts equilibrium and welfare analysis by integrating market supply and demand. Section \ref{Section Regulation} examines the effects of regulatory frictions on market equilibrium. Section \ref{Section Numerical} presents the results of numerical analysis. Finally, Section \ref{Section Conclusion} concludes the paper. 

\section{General Model and Problem Formulation} \label{Section Model} 

In this section, we build on the theoretical framework of \citet{henriet2016dynamics} and \citet{feng2023insurance}, extending it to incorporate an external capital market where insurers can invest. In this setting, the risks of insurance and financial markets are allowed to be correlated, indicating that the optimal underwriting and investment strategies must be jointly determined. As we shall see, this dependence leads to a more complex equilibrium analysis that reflects the interaction between insurance pricing and financial market conditions. 

Let $(\Omega,\mathcal{F},\{\mathcal{F}_t\}_{t\geq0},\mathbb{P})$ be a complete probability space satisfying the usual conditions. All stochastic processes governing the insurance and financial markets are assumed to be well-defined on this probability space. 

\subsection{Insurance Market} 

Consider a competitive insurance market that consists of a continuum of insurers, each offering insurance to individuals exposed to perfectly correlated risks. The idiosyncratic risks of individuals are ignored, as they can be diversified away. The cumulative loss process for a representative insurer, denoted by $L \triangleq \{ L_t: t \geq 0 \}$, evolves as the following dynamics:
\begin{equation}
    \mathrm{d}L_t = l \mathrm{d}t - \eta \mathrm{d}W^I_t, \notag
\end{equation}
where $l > 0$ represents the expected instantaneous loss rate, $\eta > 0$ denotes the exposure level to the systematic loss risk, and $W^I \triangleq \{ W^I_t: t \geq 0 \}$ is a one-dimensional standard Brownian motion modeling the risk.\footnote{In principle, the systematic loss risk may stem from multiple sources and can be modeled as multi-dimensional Brownian motion. For simplicity, and under the assumption of homogeneity in insurers' risk exposures, we aggregate them into a single dimension for tractability.} 

Insurance contracts are typically short-term,\footnote{Hence, this theoretical framework is particularly well-suited for analyzing casualty and property insurance markets, where contracts are generally short-term.} and the premium charged per unit of time is determined by the expected value principle: 
\begin{equation}
    P_t = \left(1 + \theta_t\right) \mathbb{E} [L_t] =  \left(1 + \theta_t\right) l, \notag
\end{equation}
where the premium consists of two components: an actuarially fair premium $l$, and a loading premium determined by a loading factor $\theta_t$. For simplicity, we refer to $\theta_t$ as the price of insurance. Unlike in traditional models, where $\theta_t$ is typically exogenous and non-negative, here it is endogenously determined through the equilibrium clearing of market demand and supply. By allowing $\theta_t$ to be a function of market conditions, we could characterize both the properties of $\theta_t$ and its dynamic evolution over time. 

Specifically, on the supply side, insurers determine their underwriting amount based on the prevailing insurance price. The underwriting amount at time $t$, denoted by $x_t \geq 0$, constitutes the underwriting process $X \triangleq \{ x_t: t \geq 0 \}$. Then the insurer's surplus process $ M=\{m_t: t\geq 0 \}$ evolves as the following dynamics: 
\begin{equation}
    \mathrm{d}m_t = x_t P_t \mathrm{d}t - x_t \mathrm{d}L_t = x_t \theta_t l \mathrm{d}t + x_t \eta \mathrm{d} W^I_t,  \notag
\end{equation}
where the insurer earns premium income proportional to the underwriting amount $x_t P_t \mathrm{d}t$, and is exposed to stochastic losses $x_t \mathrm{d}L_t$. 

On the demand side, the total market demand for insurance is modeled as an exogenously determined function of the loading factor $\theta_t$. The demand function, denoted by $d(\theta)$,  is continuously differentiable and satisfies: 
\begin{equation}
    d(\underline{\theta}) = 1, \quad d(\overline{\theta}) = 0, \quad d^{\prime}(\theta) < 0, \quad -1 \leq \underline{\theta} \leq \theta \leq \overline{\theta} \leq \infty. \notag
\end{equation}
Here, $\overline{\theta}$ and $\underline{\theta}$ can be interpreted as the upper and lower bounds of the loading premium, set by the regulator to ensure the market price of insurance remains within a feasible range. We also assume $\overline{\theta} > 0$ to simplify our analysis. 

In equilibrium, the market clears when the total supply of insurance equals the total demand, i.e., 
\begin{equation}
    x_t(\theta_t) = d(\theta_t), \notag
\end{equation}
and the equilibrium price $\theta^{\ast}_t$ is solved endogenously. The price process is denoted by $\Theta \triangleq \{ \theta_t: t \geq 0 \}$, while the demand process is denoted by $D \triangleq \{ d_t = d(\theta_t): t \geq 0 \}$. 

\subsection{Financial Market}

Now, assume that insurers can invest in the capital market alongside their underwriting activities. There is a classical Black-Scholes financial market in which a risk-free asset (such as a bond) and a risky asset (such as a stock or market portfolio) are traded. The dynamics of the asset prices are given by the following stochastic differential equations (SDEs): 
\begin{equation}
    \left\{
    \begin{aligned}
        & \frac{\mathrm{d}B_t}{B_t} = r \mathrm{d}t, \\
        & \frac{\mathrm{d}S_t}{S_t} = \mu \mathrm{d}t + \sigma \mathrm{d}W^S_t, 
    \end{aligned}
    \right. \notag
\end{equation}
where $r$ is the risk-free rate, $\mu > r$ represents the expected return on the risky asset, and $\sigma > 0$ denotes the volatility of the return. Here, $W^S \triangleq \{ W^S_t: t \geq 0 \}$ is a one-dimensional standard Brownian motion, which may be correlated with the insurance loss risk $W^I$. The correlation coefficient is given by $\rho \in (-1, 1)$, such that $\mathrm{d}W^I_t \mathrm{d}W^S_t = \rho \mathrm{d}t$. Here we exclude the possibility of $\rho = \pm 1$ to simplify our analysis. In practice, perfect correlation between insurance and financial risks is almost impossible, as that would imply that one of the two markets would lose its distinct purpose. But we can still analyze the implications of such extreme cases from a theoretical perspective, see Subsection \ref{Subsection Perfect Correlation}. 

In this setup, the financial market is assumed to be frictionless, implying that there are no transaction costs, taxes, or other trading constraints, and that assets are infinitely divisible. This assumption simplifies the analysis and allows insurers to freely adjust their portfolios based on market conditions. In Section \ref{Section Regulation}, we will relax this assumption and introduce regulatory frictions in the form of investment costs. 

Suppose that the insurer follows an investment strategy $Y \triangleq \{ y_t: t \geq 0 \}$, putting $y_t$ amount in the risky asset at time $t$, and the remaining $m_t - y_t$ in the risk-free asset. As a result, the insurer's wealth process $M \triangleq \{ m_t: t \geq 0 \}$ follows: 
\begin{equation}
    \mathrm{d}m_t = ( x_t \theta_t l + y_t (\mu - r) + m_t r ) \mathrm{d}t + x_t \eta \mathrm{d}W^I_t + y_t \sigma \mathrm{d} W^S_t. \label{Wealth Process}
\end{equation} 
This framework enables a detailed examination of the insurer's joint decision-making process for both underwriting and investment strategies, taking into account the interplay between insurance and financial markets. 

\subsection{Optimization Objective}

Assume that the representative insurer has a utility function $U(m)$, which satisfies $U^{\prime}(m) > 0$ and $U^{\prime \prime}(m) < 0$, reflecting risk aversion. To formalize the insurer's decision problem, we first define what constitutes an admissible strategy for the insurer. 

\begin{Definition}
    For each $t \in [0, T]$, $a \triangleq \{ a_s = (x_s, y_s): t \leq s \leq T \}$ is an admissible strategy for the insurer with initial wealth $m_t = m$ and denoted as $a \in \mathcal{A}_t$ if it satisfies: 
    \begin{enumerate}
        \item $a$ is adapted with respect to the filtration $\{\mathcal{F}_t\}_{s \in [t, T]}$; 
        \item For each $s \in [t, T]$, $x_s \geq 0$;  
        \item For any $(t, m) \in [0, T] \times \mathbb{R}$, $\mathbb{E}_{t, m} [\sup\limits_{t \leq s \leq T} \left| m_s \right|^2 ] < \infty$;
        \item SDE \eqref{Wealth Process} has a unique solution under the chosen strategy $a$. 
    \end{enumerate} 
\end{Definition} 

Given the initial wealth $m$ at time $t$, the insurer's objective is to choose an admissible strategy $a \in \mathcal{A}_t$ to maximize the expected utility of terminal wealth. The optimization problem is formalized as follows: 
\begin{equation}
    V(t, m) \triangleq \sup_{a \in \mathcal{A}_t} \mathbb{E}_{t,x} [ U(m(T)) ], \label{Objective}
\end{equation}
where $V(\cdot, \cdot)$ denotes the value function, with the boundary condition $V(T, m) = U(m)$. Based on this optimization problem, we now provide the definition of the equilibrium of the insurance market. 

\begin{Definition} \label{Equilibrium Definition}
    A stationary Markovian competitive equilibrium consists of: an insurance price process $\Theta$, an insurance demand process $D$, an insurer's wealth process $M$, an insurance supply process $X$, and an insurer's investment process $Y$, that are compatible with the insurer's optimization problem \eqref{Objective} and the market clearing condition $D = X$. 
\end{Definition} 

\section{Equilibrium Analysis} \label{Section Equilibrium}

In this section, we derive the insurer’s optimal underwriting and investment strategies by solving the stochastic control problem that incorporates the correlation between insurance and financial market risks. We then perform a supply-demand equilibrium analysis, where the equilibrium insurance price is endogenously determined by market interactions. Finally, we examine the welfare implications, focusing on how the financial market influences the overall welfare of the insurance market. 

\subsection{Solution to Decision Problem}

We begin by solving the stochastic control problem via the Hamilton-Jacobi-Bellman (HJB) equation: 
\begin{equation}
    V_{t} + \sup_{x \geq 0, y} \left\{ V_{m} \left( x \theta l  + y (\mu - r) + mr \right) + \frac{1}{2} V_{mm} ( x^2 \eta^2 + 2 \rho x \eta y \sigma + y^2 \sigma^2 ) \right\} = 0.  \label{HJB Equation}
\end{equation}
Assume that $V$ is continuously differentiable, $V_m > 0$ and $V_{mm} < 0$. The first-order conditions with respect to $x$ and $y$ are derived: 
\begin{equation}
    \left\{
    \begin{aligned}
        & x + \frac{\rho  \sigma}{\eta} y = -\frac{V_m}{V_{mm}} \frac{\theta l}{\eta^2} - \frac{1}{V_{mm}} \frac{\lambda}{\eta^2}, \\
        & y + \frac{\rho \eta}{\sigma} x = -\frac{V_m}{V_{mm}} \frac{\mu - r}{\sigma^2}, 
    \end{aligned}
    \right. \notag 
\end{equation}
where $\lambda \geq 0$ is the Lagrangian multiplier associated with the constraint $x \geq 0$. The Kuhn-Tucker condition indicates that $\lambda^{\ast} x^{\ast} = 0$, which leads to two cases. 

Case 1: $x^{\ast} = 0$. When the insurer opts not to underwrite, the investment strategy and the Lagrangian multiplier are given by
\begin{equation}
    y^{\ast} = -\frac{V_m}{V_{mm}} \frac{\mu - r}{\sigma^2}, \quad \lambda^{\ast} = V_m \left(\rho \frac{(\mu - r) \eta}{\sigma} - \theta l \right). \notag 
\end{equation}
For $\lambda^{\ast} \geq 0$, it requires that $\rho \geq \frac{\theta l \sigma}{(\mu - r) \eta}$. Substituting this result back into the HJB equation (\ref{HJB Equation}), we obtain 
\begin{equation}
    V_t + V_m mr - \frac{V_m^2}{V_{mm}} \frac{ (\mu - r)^2}{2 \sigma^2} = 0, \notag
\end{equation}
from which the value function $V$ can be solved. 

Case 2: $x^{\ast} \neq 0$. When the insurer chooses to underwrite, the Lagrange multiplier $\lambda^{\ast} = 0$, and solving the system of equations yields
\begin{equation}
    x^{\ast} = -\frac{V_m}{V_{mm}} \frac{\theta l \sigma - \rho (\mu - r) \eta}{(1 - \rho^2) \eta^2 \sigma}, \quad y^{\ast} = -\frac{V_m}{V_{mm}} \frac{(\mu - r) \eta - \rho \theta l \sigma}{(1 - \rho^2) \eta \sigma^2}. \notag    
\end{equation}
For $x^{\ast} > 0$, we require $\rho < \frac{\theta l \sigma}{(\mu - r) \eta}$. Replacing them back into the HJB equation (\ref{HJB Equation}) gives 
\begin{equation}
    V_t + V_m mr - \frac{V_m^2}{V_{mm}} \frac{ \theta^2 l^2 \sigma^2 - 2 \rho \theta l (\mu - r) \eta \sigma + (\mu - r)^2 \eta^2}{2 (1 - \rho^2) \eta^2 \sigma^2} = 0,  \notag 
\end{equation}
then the value function $V$ can be derived. 

For simplicity, we denote $\phi = \frac{\theta l \sigma}{(\mu - r)\eta}$. Economically, while $\frac{\mu - r}{\sigma}$ represents the excess return per unit of financial risk, $\frac{\theta l}{\eta}$ can be viewed as the profit per unit of insurance loss risk. Thus, $\phi$ captures the ratio of profit margins between underwriting and investment. It indicates the trade-off insurers face when allocating capital between these two activities. Based on the analysis above, the relationship between $\rho$ and $\phi$ determines the insurer's underwriting behavior, summarized in the following proposition. 

\begin{Proposition}
    Define $f(t, m, x, y) = V_{m} \left( x \theta l  + y (\mu - r) + mr \right) + \frac{1}{2} V_{mm} \left( x^2 \eta^2 + 2 \rho x \eta y \sigma + y^2 \sigma^2 \right)$. Assume  $V_m > 0$ and $V_{mm} < 0$ and that  $(x^{\ast}, y^{\ast})$ is defined as follows: 
    \begin{enumerate}
        \item if $\phi > \rho$, then 
        \begin{equation}
            (x^{\ast}, y^{\ast}) = \left(-\frac{V_m}{V_{mm}} \frac{\theta l \sigma - \rho (\mu - r) \eta}{(1 - \rho^2) \eta^2 \sigma}, -\frac{V_m}{V_{mm}} \frac{(\mu - r) \eta - \rho \theta l \sigma}{(1 - \rho^2) \eta \sigma^2}  \right). \notag 
        \end{equation} 
        \item if $\phi \leq \rho$, then 
        \begin{equation}
            (x^{\ast}, y^{\ast}) = \left(0, -\frac{V_m}{V_{mm}} \frac{\mu - r}{\sigma^2}  \right). \notag 
        \end{equation} 
    \end{enumerate}
    Then $f(t, m, x^{\ast}, y^{\ast}) = \sup\limits_{x \geq 0, y} f(t, m, x, y)$.\\  Consequently, if $a^{\ast} = \left\{ a^{\ast}_s = (x^{\ast}_s, y^{\ast}_s): (x^{\ast}_s, y^{\ast}_s) = (x^{\ast}(t, m^\ast_s), y^{\ast}(t, m^\ast _s): t \leq s \leq T \right\} \in \mathcal{A}_t$ and the HJB equation \eqref{HJB Equation} admits a solution $V$, then $a^{\ast}$ is the optimal control of Problem \eqref{Objective}, where $M^\ast = \{ m^\ast _s: s\geq t  \}$ is the unique solution of SDE \eqref{Wealth Process} when the process $a$ is replaced with $a^\ast$.
\end{Proposition} 

When $\phi > \rho$, underwriting remains competitive relative to investment, allowing insurers to maintain a presence in both activities. Because the correlation is not high enough, this diversified approach enables insurers to hedge their portfolios by generating returns from two distinct sources of risk. But when $\phi \leq \rho$, the profitability of underwriting diminishes relative to investment, prompting insurers to shift their capital entirely toward the financial market. 

\subsection{Case of CARA Utility} 

Now, we consider a specific case in which the insurer has an exponential (constant absolute risk aversion, CARA) utility: 
\begin{equation}
    U(m) = - \frac{1}{\gamma} e^{-\gamma m}, \notag 
\end{equation}
where $\gamma > 0$ represents the risk aversion level. This form of utility is widely used in financial economics because of its mathematical tractability. 

\begin{Proposition} \label{Proposition CARA Optimal Strategy}
    Given the price process $\Theta$, for an insurer with CARA utility and initial wealth $m$ at time $t$, the optimal underwriting and investment strategy is $a^{\ast} = \left\{ a^{\ast}_s = (x^{\ast}_s, y^{\ast}_s): t \leq s \leq T \right\}$, and the corresponding value function $V(t, m)$ is
    \begin{equation}
        V(t, m) = -\frac{1}{\gamma} \exp \left\{ - \gamma m e^{r(T-t)} - \int_t^T c_s(\rho, \phi_s) \mathrm{d}s \right\}, \label{Value Function}
    \end{equation}
    where $a^{\ast}_s$ and $c_s(\rho, \phi_s)$ are given as follows:
    \begin{enumerate}
        \item if $\phi_s > \rho$, then 
        \begin{eqnarray}
             &&a^{\ast}_s = \left(\frac{\theta_s l \sigma - \rho (\mu - r) \eta}{\gamma (1 - \rho^2) \eta^2 \sigma} e^{-r(T-s)} , \frac{(\mu - r) \eta - \rho \theta_s l \sigma}{\gamma (1 - \rho^2) \eta \sigma^2} e^{-r(T-s)} \right),\\  \label{Positive Underwriting}
        &&c_s(\rho, \phi_s) = \frac{ \theta_s^2 l^2 \sigma^2 - 2 \rho \theta_s l (\mu - r) \eta \sigma + (\mu - r)^2 \eta^2}{2 (1 - \rho^2) \eta^2 \sigma^2}. \notag 
        \end{eqnarray}
        \item if $\phi_s \leq \rho$, then
        \begin{equation}
            a^{\ast}_s = \left(0, \frac{\mu - r}{\gamma \sigma^2} e^{-r(T-s)} \right), \quad c_s(\rho, \phi_s) = \frac{ (\mu - r)^2}{2 \sigma^2}. \label{Zero Underwriting}
        \end{equation}
    \end{enumerate}
\end{Proposition} 
\noindent 
\begin{proof}[\textbf{Proof}.] Following \citet{yang2005optimal}, we guess that the value function takes the form of \eqref{Value Function}. It can be efficiently verified that $a^{\ast}$ is admissible and $V$ satisfies the HJB equation \eqref{HJB Equation}. Therefore, they solve  Problem (\ref{Objective}). 
\end{proof}

Note that now the price $\theta_s$ is assumed to be given, and each insurer in the competitive market acts as a price taker. Although the equilibrium price would be eventually determined through the interaction of market supply and demand, the supply function $x^{\ast}_s(\theta_s)$, as shown in \eqref{Positive Underwriting} and \eqref{Zero Underwriting}, is deterministic at each point in time, and the demand function $d(\theta_s)$ is also deterministic. As a result, the equilibrium price is not stochastic and varies only with time $s$. Insurers hold deterministic beliefs about this equilibrium price process. 

\subsection{Market Equilibrium} \label{Subsection Market Equilibrium}

Now we analyze the equilibrium conditions in the insurance market. In equilibrium, the competitive price of insurance must clear the market, meaning that the aggregate supply of insurance meets the aggregate demand, i.e., the equilibrium price $\theta^{\ast}_t$ should satisfy  the relationship(or market equilibrium equation): 
\begin{equation}
    x^{\ast}_s(\theta^{\ast}_s) = d(\theta^{\ast}_s). \notag
\end{equation}
The supply function has been given in Proposition \ref{Proposition CARA Optimal Strategy}. For simplicity, we assume that the demand function $d(\theta)$ is linear in the price $\theta$:
\begin{equation}
    d(\theta) = \frac{\overline{\theta}-\theta}{\Delta \theta}, \notag
\end{equation}
where $\Delta \theta = \overline{\theta} - \underline{\theta}$ is the difference between the upper and lower bounds of the loading premium. The size of this gap determines the elasticity of consumer demand for insurance with respect to price changes. 

Combining the demand and supply curves, we summarize the market equilibrium conditions in the following theorem. 

\begin{Theorem}[Market Equilibrium] \label{Theorem Market Equilibrium}
    Depending on parameters of insurance and financial markets, the competitive equilibrium of the insurance market is given as follows:  
    \begin{enumerate}
        \item \textbf{Positive Insurance Market}: If 
        \begin{equation}
            \overline{\theta} > \rho \frac{\mu - r}{\sigma}  \frac{\eta }{l}, \quad \underline{\theta} \leq \gamma (1 - \rho^2) \frac{\eta^2}{l} + \rho \frac{\mu - r}{\sigma}  \frac{\eta }{l}, \label{Condition Positive}
        \end{equation}
        then there is an equilibrium price process  
        \begin{equation}
            \Theta^{\ast}_p = \left\{ \theta^{\ast}_s = \frac{\gamma (1 - \rho^2) \eta^2 \overline{\theta} + \rho \frac{\mu - r}{\sigma} \eta \Delta \theta e^{-r(T - s)} }{\gamma (1 - \rho^2) \eta^2 + l \Delta \theta e^{-r(T-s)}}: t \leq s \leq T \right\}.  \notag    
        \end{equation}
        At this price, the CARA insurer optimally chooses the underwriting strategy 
        \begin{equation}
            X^{\ast}_p(\Theta^{\ast}) = \left\{ x^{\ast}_s(\theta^{\ast}_s) = \frac{ \left( \overline{\theta} l - \rho \frac{\mu - r}{\sigma} \eta \right) e^{-r(T-s)} }{\gamma (1 - \rho^2) \eta^2 + l \Delta \theta e^{-r(T-s)}}:t \leq s \leq T \right\},  \notag
        \end{equation}
        which is strictly positive, and the investment strategy 
        \begin{equation}
            Y^{\ast}_p(\Theta^{\ast}) = \left\{ y^{\ast}_s(\theta^{\ast}_s) = \frac{ \left[ \frac{\eta^2}{\sigma} \left( \frac{\mu-r}{\sigma} - \rho \frac{l}{\eta} \overline{\theta} \right) + \frac{\mu-r}{\gamma \sigma^2} l \Delta \theta e^{-r(T-s)} \right] e^{-r(T-s)} }{\gamma (1 - \rho^2) \eta^2 + l \Delta \theta e^{-r(T-s)}}: t \leq s \leq T \right\}.  \notag
        \end{equation}
        \item \textbf{Zero Insurance Market}: If 
        \begin{equation}
            \overline{\theta} \leq \rho \frac{\mu - r}{\sigma}  \frac{\eta }{l}, \label{Condition Zero}
        \end{equation}
        then there is an equilibrium price process  
        \begin{equation}
            \Theta^{\ast}_z = \left\{ \theta^{\ast}_s = \overline{\theta}: t \leq s \leq T \right\}. \notag     
        \end{equation}
        At this price, the CARA insurer optimally chooses the underwriting strategy 
        \begin{equation}
            X^{\ast}_z(\Theta^{\ast}) = \left\{ x^{\ast}_s(\theta^{\ast}_s) = 0:  t \leq s \leq T \right\}, \notag 
        \end{equation}
        which keeps zero, and the investment strategy 
        \begin{equation}
            Y^{\ast}_z(\Theta^{\ast}) = \left\{ y^{\ast}_s(\theta^{\ast}_s) = \frac{\mu - r}{\gamma \sigma^2} e^{-r(T-s)}: t \leq s \leq T \right\}. \notag  
        \end{equation} 
        \item \textbf{Market Failure}: If 
        \begin{equation}
            \overline{\theta} > \rho \frac{\mu - r}{\sigma} \frac{\eta}{l}, \quad \underline{\theta} > \gamma (1 - \rho^2) \frac{\eta^2}{l} + \rho \frac{\mu - r}{\sigma}  \frac{\eta }{l}, \label{Condition Failure}
        \end{equation}
        then there exists a stopping time $\tau_f = \inf\limits_{s \geq t} \{ \underline{\theta} > \gamma (1 - \rho^2) \frac{\eta^2}{l}e^{r(T-s)} + \rho \frac{\mu - r}{\sigma}  \frac{\eta }{l} \}$. When $\tau_f \leq s \leq T$, the market fails. Thus, there is no equilibrium in the insurance market. 
    \end{enumerate} 
\end{Theorem} 
\noindent
\begin{proof}[\textbf{Proof}.] Based on Proposition \ref{Proposition CARA Optimal Strategy}, when $\phi_s > \rho$, the insurer actively participates in both underwriting and investment. In this case, the supply function is given by $x^{\ast}_s(\theta_s) = \frac{\theta_s l \sigma - \rho (\mu - r) \eta}{\gamma (1 - \rho^2) \eta^2 \sigma} e^{-r(T-s)}$, which is an upward-sloping linear curve and has an intersection with the downward-sloping demand curve. But to further guarantee the intersection is an equilibrium, we need 
\begin{equation}
    \underline{\theta} \leq \theta^{\ast}_s \leq \overline{\theta}, \quad \phi_s(\theta^{\ast}_s) > \rho, \notag    
\end{equation}
from which conditions on $\overline{\theta}$ and $\underline{\theta}$ can be derived, as shown in \eqref{Condition Positive}.  

When $\phi_s \leq \rho$, the insurer stops underwriting, and $x^{\ast}_s = 0$. The equilibrium price becomes $\theta^{\ast}_s = \overline{\theta}$, and it should satisfy $\phi_s(\theta^{\ast}_s) \leq \rho$. Then Condition \eqref{Condition Zero} is derived. Beyond these two cases, when $\underline{\theta}$ is set relatively high, the intersection would violate $\underline{\theta} \leq \theta^{\ast}_s$, meaning there is no feasible equilibrium. This results in the condition specified in \eqref{Condition Failure}. 
\end{proof}

\begin{figure}[htb]
    \centering
    \includegraphics[width=1.0\linewidth]{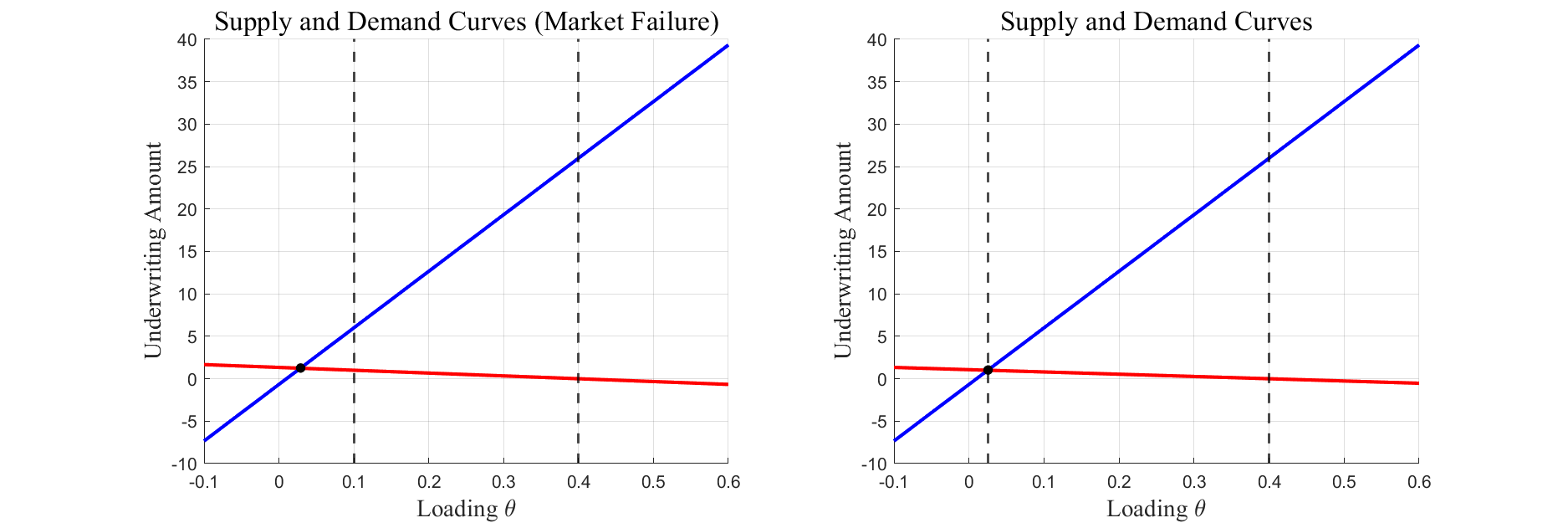}
    \caption{Case of Market Failure}
    \label{case failure}
\end{figure}

As discussed, $\phi = \frac{\theta l \sigma}{(\mu - r)\eta}$ reflects the profit margin ratio between underwriting and investment. For conditions in Theorem \ref{Theorem Market Equilibrium}, $\frac{\overline{\theta} l \sigma}{(\mu - r)\eta}$ represents the feasible maximum profit margin ratio. A necessary condition for the existence of a positive insurance market is that this maximum ratio exceeds the market correlation $\rho$. Otherwise, underwriting would become unprofitable, leading insurers to give up underwriting. The other condition $\underline{\theta} \leq \gamma (1 - \rho^2) \frac{\eta^2}{l} + \rho \frac{\mu - r}{\sigma}  \frac{\eta }{l}$, imposes a restriction on the lower bound of the loading. If this condition is violated, the market fails, as illustrated in Figure \ref{case failure}. The supply and demand curves demonstrate that even at the minimum premium, the quantity of insurance supplied can exceed the quantity demanded, resulting in a market imbalance where there is no equilibrium price to clear the market. This result suggests that regulators should avoid setting the lower bound of insurance prices too high, as insurers may reduce price to struggle to remain competitive, ultimately leading to market failure and a loss of social welfare.\footnote{In practice, setting insurance prices too low may expose insurers to solvency risks, especially during unexpected losses. This is why regulators impose strict restrictions on insurance pricing. While we assume underwriting and investment are managed within a single account in this paper, real-world regulations often treat solvency and investment practices separately. } 

\begin{Proposition} \label{Proposition Comparative Static}
    Under the positive insurance market condition \eqref{Condition Positive}: 
    \begin{enumerate}
        \item $\frac{\partial \theta^{\ast}_s}{\partial \left(\frac{\mu-r}{\sigma}\right)} < 0$, $= 0$ or $> 0$, if and only if $\rho < 0$, $= 0$ or $> 0$. The equilibrium insurance price decreases (increases) with the investment Sharpe ratio, when the two market risks are negatively (positively) correlated. 
        \item $\frac{\partial x^{\ast}_s}{\partial \left(\frac{\mu-r}{\sigma}\right)} < 0$, $= 0$ or $> 0$, if and only if $\rho > 0$, $= 0$ or $< 0$. The equilibrium underwriting amount decreases (increases) with the investment Sharpe ratio, when the two market risks are positively (negatively) correlated. 
        \item $\frac{\partial y^{\ast}_s}{\partial \left(\frac{\mu-r}{\sigma}\right)} > 0$. The equilibrium investment amount always increases with the investment Sharpe ratio. 
    \end{enumerate}
\end{Proposition}

While \citet{knox2024insurers} theoretically and empirically demonstrate that insurers would set lower prices when expected investment returns are higher, results in Proposition \ref{Proposition Comparative Static} refine this conclusion by restricting it to the condition $\rho < 0$. When the correlation between insurance losses and financial returns is negative, insurers benefit from diversification, enabling them to reduce premiums. However, the proposition also reveals that this relationship can reverse when $\rho > 0$. In the case of positive correlation, insurers may actually raise prices to account for the compounded risks in both markets. While \citet{knox2024insurers} focus on returns from illiquid investments, our analysis highlights the crucial role of the correlation coefficient in determining the direction of pricing adjustments. 

\subsection{Case of Zero Correlation} \label{Subsection Zero Correlation}

To further understand the equilibrium loading premium, we now explore two specific cases. The first case assumes no correlation between insurance and financial market risks ($\rho = 0$), which simplifies the interaction between underwriting and investment. 

\begin{Proposition}
    Given $\rho = 0$, and the positive insurance market condition \eqref{Condition Positive}, the equilibrium price process is $\Theta^{\ast} = \left\{ \theta^{\ast}_s = \frac{\gamma \eta^2 \overline{\theta}}{\gamma \eta^2 + l \Delta \theta e^{-r(T-s)}}: t \leq s \leq T \right\}$, which is strictly positive. 
\end{Proposition}

It can be verified that this equilibrium loading  is equivalent to the scenario where there is no investment opportunity, indicating that underwriting and investment strategies are independent when the two risks are uncorrelated. The strictly positive loading indicates that insurers should always realize positive expected profits from their insurance operations, consistent with the finding of \citet{henriet2016dynamics}. 

Additionally, the loading increases with risk aversion ($\gamma$) and volatility ($\eta$), as more risk-averse insurers or those facing higher volatility require higher premiums to offset uncertainty. But it decreases with expected loss ($l$), because  higher loss expectations require balancing competitive pricing. If the expected loss is too high, increasing the loading could reduce demand and result in lower price. The loading also declines with a wider pricing range ($\Delta \theta$), as greater flexibility would intensify competition between insurers and therefore drive down the loading. Finally, the loading decreases over time ($s$), reflecting diminished risk exposure as the insurance term progresses. 

\subsection{Case of Negative Loading} \label{Subsection Negative Loading} 

While the loading is positive when the risks are uncorrelated, it can become negative when there is a relatively strong negative correlation between insurance losses and financial market returns. In such cases, insurers may lower premiums to attract policyholders and invest the collected premiums in higher-yielding assets. 

\begin{Proposition} \label{Proposition Negative Loading}
    Given $\frac{\rho}{1 - \rho^2} < - \gamma \frac{\sigma}{\mu - r} \frac{\eta \overline{\theta}}{\Delta \theta}$, and the positive insurance market condition \eqref{Condition Positive}, there exists a stopping time $\tau_n = \inf\limits_{s \geq t} \left\{ \frac{\rho}{1 - \rho^2} < - \gamma \frac{\sigma}{\mu - r} \frac{\eta \overline{\theta}}{\Delta \theta} e^{r(T-s)} \right\}$. When $\tau_n \leq s \leq T$, the equilibrium loading premium $\theta^{\ast}_s$ becomes negative. 
\end{Proposition}

Clearly, the condition for a negative loading requires $\rho < 0$, and it can be easily achieved as $\rho$ approaches $-1$. In such cases, insurers are incentivized to offer negative premiums, effectively subsidizing policyholders, because the expected investment returns from the financial market exceed potential underwriting losses. This approach reflects the financialization of modern insurers, who leverage premiums as investment capital rather than solely focusing on underwriting profits. 

The likelihood of negative loading increases under favorable market conditions. Lower risk aversion ($\gamma$) and volatility of insurance losses ($\eta$) make insurers more willing to accept underwriting losses for the prospect of higher investment returns. A higher Sharpe ratio ($\frac{\mu - r}{\sigma}$) further incentivizes insurers to reduce premiums because  they anticipate more investment returns. For price regulation, a higher upper bound ($\overline{\theta}$) makes it harder to achieve negative loading, as insurers can still charge positive premiums. In contrast, a wider range of premium flexibility ($\Delta \theta$) intensifies competition and drives premiums lower. 

\subsection{Welfare Implication} \label{Welfare Implication} 

In this section, we analyze the welfare implications of the equilibrium by calculating social welfare as the sum of consumer surplus and producer surplus. Consumer surplus reflects the benefit consumers get from paying less than they are willing to, while producer surplus represents the difference between insurers' earnings and underwriting costs. Together, they provide an overall measure of market efficiency. 

\begin{Theorem}[Social Welfare] \label{Theorem Social Welfare}
    Under the positive insurance market condition \eqref{Condition Positive}, the social welfare of the insurance market is given by 
    \begin{equation}
        W^{\ast}(\Theta^{\ast}) = \left\{ w^{\ast}_s(\theta^{\ast}_s) = \frac{ \frac{1}{2} l \left( \overline{\theta} - \rho \frac{\mu - r}{\sigma} \frac{\eta}{l} \right)^2 e^{-r(T-s)} }{\gamma (1 - \rho^2) \eta^2 + l \Delta \theta e^{-r(T-s)}}: t \leq s \leq T \right\}. \notag    
    \end{equation} 
    Furthermore, $\frac{\partial w^{\ast}_s}{\partial \left(\frac{\mu-r}{\sigma}\right)} < 0$, $= 0$ or $> 0$, if and only if $\rho > 0$, $= 0$ or $< 0$. This implies that social welfare decreases (increases) with the investment Sharpe ratio when the two market risks are positively (negatively) correlated. 
\end{Theorem}

As shown, the performance of financial market affects the overall welfare of insurance market when the two markets are correlated. 

\subsection{Case of Perfect Correlation} \label{Subsection Perfect Correlation}

Although perfect correlation between insurance and financial market risks is rare in practice, we analyze the equilibrium outcomes in this subsection. When $\rho = \pm 1$, the maximization condition in \eqref{HJB Equation} exists if and only if $\frac{\theta l}{\eta} = \pm \frac{\mu - r}{\sigma}$. Then the equilibrium price $\theta^{\ast}$ is derived and we have the following. 

\begin{Proposition} \label{Proposition Perfect Correlation}
    Given $\rho = \pm 1$, and $\underline{\theta} \leq \rho \frac{\mu - r}{\sigma} \frac{\eta}{l} \leq \overline{\theta}$, there is an equilibrium price process
        \begin{equation}
            \Theta^{\ast} = \{ \theta^{\ast}_s = \rho \frac{\mu - r}{\sigma} \frac{\eta}{l}: t \leq s \leq T \}. \notag     
        \end{equation}
        At this price, the CARA insurer optimally chooses its underwriting strategy 
        \begin{equation}
            X^{\ast}(\Theta^{\ast}) = \{ x^{\ast}_s(\theta^{\ast}_s) = \frac{\overline{\theta} - \rho \frac{\mu - r}{\sigma} \frac{\eta}{l}}{\Delta \theta}: t \leq s \leq T \}, \notag 
        \end{equation}
        and its investment strategy 
        \begin{equation}
            Y^{\ast}(\Theta^{\ast}) = \{ y^{\ast}_s(\theta^{\ast}_s) = \frac{ \frac{\eta^2}{\sigma} \left( \frac{\mu-r}{\sigma} - \rho \frac{l}{\eta} \overline{\theta} \right) + \frac{\mu-r}{\gamma \sigma^2} l \Delta \theta e^{-r(T-s)} }{l \Delta \theta}: t \leq s \leq T \}. \notag  
        \end{equation}
        Results here align with those in Theorem \ref{Theorem Market Equilibrium} in terms of expression if substituting $\rho = \pm 1$.
\end{Proposition}

\section{Equilibrium with Regulatory Friction} \label{Section Regulation} 

In practice, risk-based capital regulations are designed to limit insurers' risk-shifting behavior that could arise from limited liability or state guaranty associations. These regulation often impose costs on insurers' investment activities, particularly when they take on higher levels of financial risk \citep{koijen2022fragility, koijen2023understanding}. In this section, we analyze how such constraints affect the optimal underwriting and investment strategies of insurers, as well as their overall impact on the market equilibrium. 

\subsection{Market Equilibrium}

We model the costs of regulatory frictions as quadratic functions of investment capital, reducing the insurer's net cash flow: 
\begin{equation}
    \mathrm{d}m_t = \big( x_t \theta_t l + y_t (\mu - r) - \frac{1}{2} y_t^2 \epsilon_t + m_t r \big) \mathrm{d}t + x_t \eta \mathrm{d}W^I_t + y_t \sigma \mathrm{d} W^S_t, \notag
\end{equation} 
where $\epsilon_t \in [0, \overline{\epsilon}]$ denotes the intensity of regulatory frictions and consists of $\mathcal{E} \triangleq \left\{ \epsilon_t: t \geq 0 \right\}$. And $\overline{\epsilon}$ is an exogenously given bar. Using a similar approach to the derivation in Section \ref{Section Equilibrium}, we obtain the following proposition describing the insurer's optimal strategy. 

\begin{Proposition} \label{Proposition CARA Optimal Strategy with Friction}
    Given the price process $\Theta$ and the regulatory cost process $\mathcal{E}$, for an insurer with CARA utility and initial wealth $m$ at time $t$, the optimal underwriting and investment strategy is $a^{\ast} = \left\{ a^{\ast}_s = (x^{\ast}_s, y^{\ast}_s): t \leq s \leq T \right\}$, and corresponding value function $V(t, m)$ is 
    \begin{equation}
        V(t, m) = -\frac{1}{\gamma} \exp \left\{ - \gamma m e^{r(T-t)} 
        - \int_t^T h_s(\rho, \phi_s) \mathrm{d}s \right\}, \notag  
    \end{equation}
    where $a^{\ast}_s$ and $h_s(\rho, \phi_s)$ are respectively  as follows: 
    \begin{enumerate}
        \item if $\left(1 + \frac{\epsilon_s}{\gamma \sigma^2} e^{-r(T-s)}\right) \phi_s > \rho$, then 
        \begin{eqnarray*}
            &&a^{\ast}_s = \left(\frac{\left(1 + \frac{\epsilon_s}{\gamma \sigma^2} e^{-r(T-s)}\right) \theta_s l \sigma - \rho (\mu - r) \eta}{\gamma \left(1 + \frac{\epsilon_s}{\gamma \sigma^2} e^{-r(T-s)} - \rho^2\right) \eta^2 \sigma} e^{-r(T-s)}, \frac{(\mu - r) \eta - \rho \theta_s l \sigma}{\gamma \left(1 + \frac{\epsilon_s}{\gamma \sigma^2} e^{-r(T-s)} - \rho^2\right) \eta \sigma^2} e^{-r(T-s)}\right), \\        
           && h_s(\rho, \phi_s) =\frac{ \left(1 + \frac{\epsilon_s}{\gamma \sigma^2} e^{-r(T-s)}\right) \theta_s^2 l^2 \sigma^2 
            - 2 \rho \theta_s l (\mu - r) \eta \sigma + (\mu - r)^2 \eta^2}
            {2 \left(1 + \frac{\epsilon_s}{\gamma \sigma^2} e^{-r(T-s)} - \rho^2\right) \eta^2 \sigma^2}.  
        \end{eqnarray*} 
        \item if $\left(1 + \frac{\epsilon_s}{\gamma \sigma^2} e^{-r(T-s)}\right) \phi_s \leq \rho$, then 
        \begin{equation}
            a^{\ast}_s = \left(0, \frac{\mu - r}{\gamma \sigma^2 e^{r(T-s)} + \epsilon_s} \right), \quad h_s(\rho, \phi_s) = \frac{ (\mu - r)^2}{2 \left(1 + \frac{\epsilon_s}{\gamma \sigma^2}e^{-r(T-s)} \right) \sigma^2}.  \notag
        \end{equation}
    \end{enumerate}
\end{Proposition} 

Compared with the results in Proposition \ref{Proposition CARA Optimal Strategy}, the introduction of a positive regulatory cost $\epsilon_t$ alters the insurer's optimal strategy. While the original $\phi$ denotes the ratio of profit margins between underwriting and investment activities, here it is adjusted to $\left(1 + \frac{\epsilon_s}{\gamma \sigma^2} e^{-r(T-s)}\right) \phi$, and reduces the net return from financial investments due to the added cost of holding risky assets. This adjustment may effectively incentivize the insurer to maintain or even increase their underwriting position. 

\begin{Theorem}[Market Equilibrium with Regulatory Friction] \label{Theorem Market Equilibrium with Friction}
        If for any $s \in [t, T]$, 
        \begin{equation}
            \overline{\theta} > \frac{1}{1 + \frac{\epsilon_s}{\gamma \sigma^2} e^{-r(T-s)} } \rho \frac{\mu - r}{\sigma}  \frac{\eta }{l}, \quad \underline{\theta} \leq \frac{1}{1 + \frac{\epsilon_s}{\gamma \sigma^2} e^{-r(T-s)} } \left( \gamma (1 - \rho^2) \frac{\eta^2}{l} + \rho \frac{\mu - r}{\sigma}  \frac{\eta }{l} \right), \label{Regulated Positive Market Condition}
        \end{equation}
        then there is an equilibrium price process  
        \begin{equation}
            \Theta_p^{\ast} = \left\{ \theta^{\ast}_s = \frac{\gamma \left(1 + \frac{\epsilon_s}{\gamma \sigma^2} e^{-r(T-s)} - \rho^2\right) \eta^2 \overline{\theta} + \rho \frac{\mu - r}{\sigma} \eta \Delta \theta e^{-r(T - s)} }{\gamma \left(1 + \frac{\epsilon_s}{\gamma \sigma^2}e^{-r(T-s)} - \rho^2\right) \eta^2 + \left(1 + \frac{\epsilon_s}{\gamma \sigma^2} e^{-r(T-s)} \right) l \Delta \theta e^{-r(T-s)}}: t \leq s \leq T \right\}. \notag
        \end{equation}
        At this price, the CARA insurer optimally chooses the underwriting strategy 
        \begin{equation}
            X_p^{\ast}(\Theta^{\ast}) = \left\{ x^{\ast}_s(\theta^{\ast}_s) = \frac{ \left[ \left(1 + \frac{\epsilon_s}{\gamma \sigma^2} e^{-r(T-s)} \right) \overline{\theta} l - \rho \frac{\mu - r}{\sigma} \eta \right] e^{-r(T-s)} }{\gamma \left(1 + \frac{\epsilon_s}{\gamma \sigma^2} e^{-r(T-s)} - \rho^2\right) \eta^2 + \left(1 + \frac{\epsilon_s}{\gamma \sigma^2} e^{-r(T-s)} \right) l \Delta \theta e^{-r(T-s)}}: t \leq s \leq T \right\}, \notag
        \end{equation}
        which is strictly positive, and the investment strategy 
        \begin{equation}
            Y_p^{\ast}(\Theta^{\ast}) = \left\{ y^{\ast}_s(\theta^{\ast}_s) = \frac{ \left[ \frac{\eta^2}{\sigma} \left( \frac{\mu-r}{\sigma} -  \rho \frac{l}{\eta} \overline{\theta} \right) + \frac{\mu - r}{\gamma \sigma^2} l \Delta \theta e^{-r(T-s)}  \right] e^{-r(T-s)} }{\gamma \left(1 + \frac{\epsilon_s}{\gamma \sigma^2} e^{-r(T-s)} - \rho^2\right) \eta^2 + \left(1 + \frac{\epsilon_s}{\gamma \sigma^2} e^{-r(T-s)} \right) l \Delta \theta e^{-r(T-s)}}: t \leq s \leq T \right\}. \notag
        \end{equation}      
\end{Theorem} 
To save space, here we only provide the equilibrium results for a positive insurance market, which is the case we are most interested in. This is consistent with Theorem \ref{Theorem Market Equilibrium} when we set $\epsilon_s = 0$. The cases of a zero insurance market and market failure can be similarly derived with ease. To explore the impact of regulatory frictions on market equilibrium, we conduct the following comparative static analysis. 

\begin{Proposition} \label{Proposition Regulatory Comparative Static}
    Denote $\overline{\rho}_s = \frac{1}{\overline{\theta}} \frac{\mu - r}{\sigma} \frac{\eta}{l} \left( 1 + \frac{l \Delta \theta}{\gamma \eta^2} e^{-r(T-s)} \right)$. Under the positive insurance market condition \eqref{Regulated Positive Market Condition}:  
    \begin{enumerate}
        \item $\frac{\partial \theta^{\ast}_s}{\partial \epsilon} > 0$, if and only if $\rho < 0$, or $\rho > \overline{\rho}_s$; 
        
        $\frac{\partial \theta^{\ast}_s}{\partial \epsilon} = 0$, if and only if $\rho = 0$, or $\rho = \overline{\rho}_s$; 
        
        $\frac{\partial \theta^{\ast}_s}{\partial \epsilon} < 0$, if and only if $0 < \rho < \overline{\rho}_s$. 
        
        \item $\frac{\partial x^{\ast}_s}{\partial \epsilon} < 0$, if and only if $\rho < 0$ or $\rho >\overline{\rho}_s$; 
        
        $\frac{\partial x^{\ast}_s}{\partial \epsilon} = 0$, if and only if $\rho = 0$, or $\rho = \overline{\rho}_s$; 
        
        $\frac{\partial x^{\ast}_s}{\partial \epsilon} > 0$, if and only if $0 < \rho < \overline{\rho}_s$. 
        
        \item $\frac{\partial y^{\ast}_s}{\partial \epsilon} > 0$, $= 0$ or $< 0$, if and only if $y^{\ast}_s < 0$, $= 0$ or $> 0$. 
    \end{enumerate}
\end{Proposition} 

Although we have discussed that a positive regulatory cost increases the likelihood that the insurer maintains a positive underwriting position, within a positive insurance market, the rise in regulatory cost does not necessarily lead to an increase in equilibrium underwriting. The impact of regulatory friction on equilibrium price and insurer behavior depends on the correlation coefficient. When $\rho = 0$, the insurance market equilibrium is independent of the financial market performance and frictions. However, when $\rho \neq 0$, the price of insurance is influenced by both insurance and financial markets simultaneously.

Under certain conditions, when $\rho < 0$ or $\rho > \overline{\rho}_s$, an increase in regulatory costs may raise the price of insurance, reducing market demand and thus lowering the equilibrium quantity of insurance. In the intermediate range ($0 < \rho < \overline{\rho}_s$), the two markets behave complementarily. An increase in regulatory costs would reduce the equilibrium price, thereby encouraging a higher underwriting volume. These findings highlight the complexity of the interplay between regulatory frictions and the insurance market equilibrium.  

\subsection{Welfare Implication}
In this subsection, we examine how regulatory frictions impact the social welfare of the insurance market and explore the conditions for the regulator to set an optimal cost. 

\begin{Theorem}[Social Welfare with Regulatory Friction] \label{Theorem Regulated Social Welfare}
    Under the positive insurance market condition \eqref{Regulated Positive Market Condition}, the social welfare of the insurance market is given by 
    \begin{equation}
        W^{\ast}(\Theta^{\ast}) = \left\{ w^{\ast}_s(\theta^{\ast}_s) = \frac{ \frac{1}{2} \left( 1 + \frac{\epsilon_s}{\gamma \sigma^2} e^{-r(T-s)}\right) l \left( \overline{\theta} - \frac{1}{1 + \frac{\epsilon_s}{\gamma \sigma^2}e^{-r(T-s)}} \rho \frac{\mu - r}{\sigma} \frac{\eta}{l} \right)^2 e^{-r(T-s)} }{\gamma \left(1 + \frac{\epsilon_s}{\gamma \sigma^2} e^{-r(T-s)} - \rho^2\right) \eta^2 + \left(1 + \frac{\epsilon_s}{\gamma \sigma^2} e^{-r(T-s)}\right) l \Delta \theta e^{-r(T-s)}}: t \leq s \leq T \right\}. \notag    
    \end{equation} 
    Define $\overline{\rho}_s = \frac{1}{\overline{\theta}} \frac{\mu - r}{\sigma} \frac{\eta}{l} \left( 1 + \frac{l \Delta \theta}{\gamma \eta^2} e^{-r(T-s)} \right)$, $\underline{\rho}_s = -\frac{\overline{\theta}}{2} \frac{\sigma}{\mu - r} \frac{l}{\eta} + \frac{1}{2} \sqrt{\overline{\theta}^2 \left(\frac{\sigma}{\mu - r}\right)^2 \left(\frac{l}{\eta}\right)^2 + 8 \left(1 + \frac{\Delta \theta l}{\gamma \eta^2} e^{-r(T-s)}\right)}$, $\overline{\epsilon}_s = \gamma \sigma^2 \left[\frac{1}{\underline{\theta}}\left( \gamma (1 - \rho^2) \frac{\eta^2}{l} + \rho \frac{\mu - r}{\sigma}  \frac{\eta }{l} \right) - 1 \right]e^{r(T-s)}$, and $\underline{\epsilon}_s = \gamma \sigma^2 \left(\frac{1}{\overline{\theta}} \rho \frac{\mu - r}{\sigma} \frac{\eta}{l} - 1 \right) e^{r(T-s)}$. For convenience, we assume $\overline{\epsilon}_s < 0$, and $\underline{\epsilon}_s < 0$.\footnote{This assumption does not significantly affect the results. Without the assumption $\underline{\epsilon}_s < 0$, $\epsilon^{\ast}_s = 0$ in cases (1) and (3) would be replaced by $\epsilon^{\ast}_s = \max \left\{0, \underline{\epsilon}_s\right\}$. But when $\epsilon^{\ast}_s = \underline{\epsilon}_s$, the equilibrium quantity of insurance becomes zero. And without the assumption $\overline{\epsilon}_s < 0$, $\epsilon^{\ast}_s = \overline{\epsilon}$ in cases (2) and (3) would be replaced by $\epsilon^{\ast}_s = \min \left\{\overline{\epsilon}, \overline{\epsilon}_s\right\}$. } At time $s \in [t, T]$, the social welfare is maximized at $\epsilon^{\ast}_s$, which is given as follows: 
    \begin{enumerate}
        \item if $\rho < 0$ or $\rho \geq 2 \overline{\rho}_s$, then $\epsilon^{\ast}_s = 0$. 
        \item if $0 < \rho < \underline{\rho}_s$, then $\epsilon^{\ast}_s = \overline{\epsilon}$. 
        \item if $\underline{\rho}_s \leq \rho < 2 \overline{\rho}_s$, then $\epsilon^{\ast}_s = 0$ or $\overline{\epsilon}$, whichever results in greater welfare. 
        \item if $\rho = 0$, then social welfare is independent of $\epsilon_s$. 
    \end{enumerate}
\end{Theorem}
\noindent 
\begin{proof}[\textbf{Proof}.]
The social welfare can be easily calculated by $w_s^{\ast} = \frac{1}{2} x_s^{\ast} \left(\overline{\theta} - \frac{\rho (\mu - r)\eta}{\left(1 + \frac{\epsilon_s}{\gamma \sigma^2} e^{-r(T-s)} \right) l \sigma} \right)$. Define $\alpha_s = \frac{1}{1 + \frac{\epsilon_s}{\gamma \sigma^2} e^{-r(T-s)}} \in (0, 1]$. Then the derivative of social welfare with respect to the cost is
\begin{equation}
    \frac{\partial w_s^{\ast}}{\partial \epsilon} = \beta_s \frac{\partial \alpha_s}{\partial \epsilon} \rho \left[\rho^2 \alpha_s + \rho \overline{\theta} \frac{\sigma}{\mu - r} \frac{l}{\eta} - 2 \left( 1 + \frac{l \Delta \theta}{\gamma \eta^2} e^{-r(T-s)} \right) \right], \notag
\end{equation}
where $\beta_s$ is a strictly positive function of $\epsilon_s$, and $\frac{\partial \alpha_s}{\partial \epsilon} < 0$ because it is a decreasing function of $\epsilon_s$. If $\rho = 0$, then $\frac{\partial w_s^{\ast}}{\partial \epsilon} = 0$. If $\rho < 0$, with Condition \eqref{Regulated Positive Market Condition}, we have 
\begin{equation}
    \rho^2 \alpha_s + \rho \overline{\theta} \frac{\sigma}{\mu - r} \frac{l}{\eta} - 2 \left( 1 + \frac{l \Delta \theta}{\gamma \eta^2} e^{-r(T-s)} \right) < 2 \rho^2 \alpha_s - 2 < 0. \notag
\end{equation}
Thus $\frac{\partial w_s^{\ast}}{\partial \epsilon} < 0$. If $\rho > 0$, setting $\frac{\partial w_s^{\ast}}{\partial \epsilon} = 0$ yields
\begin{equation}
    \alpha_s = \frac{1}{\rho^2} \left[ 2 \left( 1 + \frac{l \Delta \theta}{\gamma \eta^2} e^{-r(T-s)} \right) - \rho \overline{\theta} \frac{\sigma}{\mu - r} \frac{l}{\eta} \right] \in (0, 1], \notag
\end{equation}
which requires $\underline{\rho}_s \leq \rho < 2 \overline{\rho}_s$. If $0 < \rho < \underline{\rho}_s$, the term inside the bracket is negative and we always have $\frac{\partial w_s^{\ast}}{\partial \epsilon} > 0$. Last, if $\rho \geq 2 \overline{\rho}_s$, the term inside the bracket becomes positive and we always have $\frac{\partial w_s^{\ast}}{\partial \epsilon} < 0$. 

Condition \eqref{Regulated Positive Market Condition} restricts that $\epsilon_s \in (\underline{\epsilon}_s, \overline{\epsilon}_s]$ if $\underline{\theta} > 0$. And $\epsilon_s > \underline{\epsilon}_s$, $\epsilon_s \geq \overline{\epsilon}_s$ if $\underline{\theta} < 0$. As we assume that $\overline{\epsilon}_s < 0$, and $\underline{\epsilon}_s < 0$, feasible $\epsilon_s$ ranges from $0$ to $\overline{\epsilon}$. If $\rho < 0$ or $\rho \geq 2 \overline{\rho}_s$, the welfare always decreases in $\epsilon_s$, therefore the optimal cost is $\epsilon^{\ast}_s = 0$. If $0 < \rho < \underline{\rho}_s$, the welfare always increases in $\epsilon_s$, then the optimal cost is $\epsilon^{\ast}_s = \overline{\epsilon}$. If $\underline{\rho}_s \leq \rho < 2 \overline{\rho}_s$, the welfare decreases in $\epsilon_s$ first and them increases in it, reaching the maximum at $0$ or $\overline{\epsilon}$.  
\end{proof}

An important implication of Theorem \ref{Theorem Regulated Social Welfare} is that the intensity of regulation that maximizes social welfare in the insurance market can be either zero or a positive value, depending on the correlation coefficient between insurance and financial markets. When $\rho$ is less than zero or a significantly large positive number, as discussed, an increase in regulatory costs raises the price of insurance, reduces market demand and ultimately leads to a loss in social welfare. However, in the intermediate range of $\rho$, an admissible positive regulatory cost may actually improve social welfare. And in this case, the optimal regulatory cost leads to an equilibrium price that reaches $\underline{\theta}$, which is the lowest price allowed by the regulator. Therefore, understanding the magnitude of $\rho$ in real markets is crucial for setting the optimal regulatory policy. Even if it is difficult to measure its exact value, knowing the potential range of $\rho$ can still provide valuable guidance in determining the appropriate regulatory approach.

\section{Numerical Analysis} \label{Section Numerical}

In this section, we provide a numerical illustration of the market equilibrium. For the benchmark case, we set parameter values as follows: 
\begin{table}[htbp]
\centering
\setlength{\tabcolsep}{12.5pt}
\renewcommand{\arraystretch}{1.2}
\caption{Benchmark Parameters}
\small 
\begin{tabular}{lcccccccccc}
\hline
Parameter & $l$   & $\eta$ & $r$   & $\mu$  & $\sigma$ & $\gamma$ & $t$  & $T$ & $\overline{\theta}$ & $\underline{\theta}$ \\ \hline
Value     & 0.5   & 0.1    & 0.04  & 0.12   & 0.2      & 2.0      & 0    & 50 & 0.2 & -0.2 \\ \hline
\end{tabular}
\end{table}

\begin{figure}[htpb]
    \centering
    \includegraphics[width=1.0\linewidth]{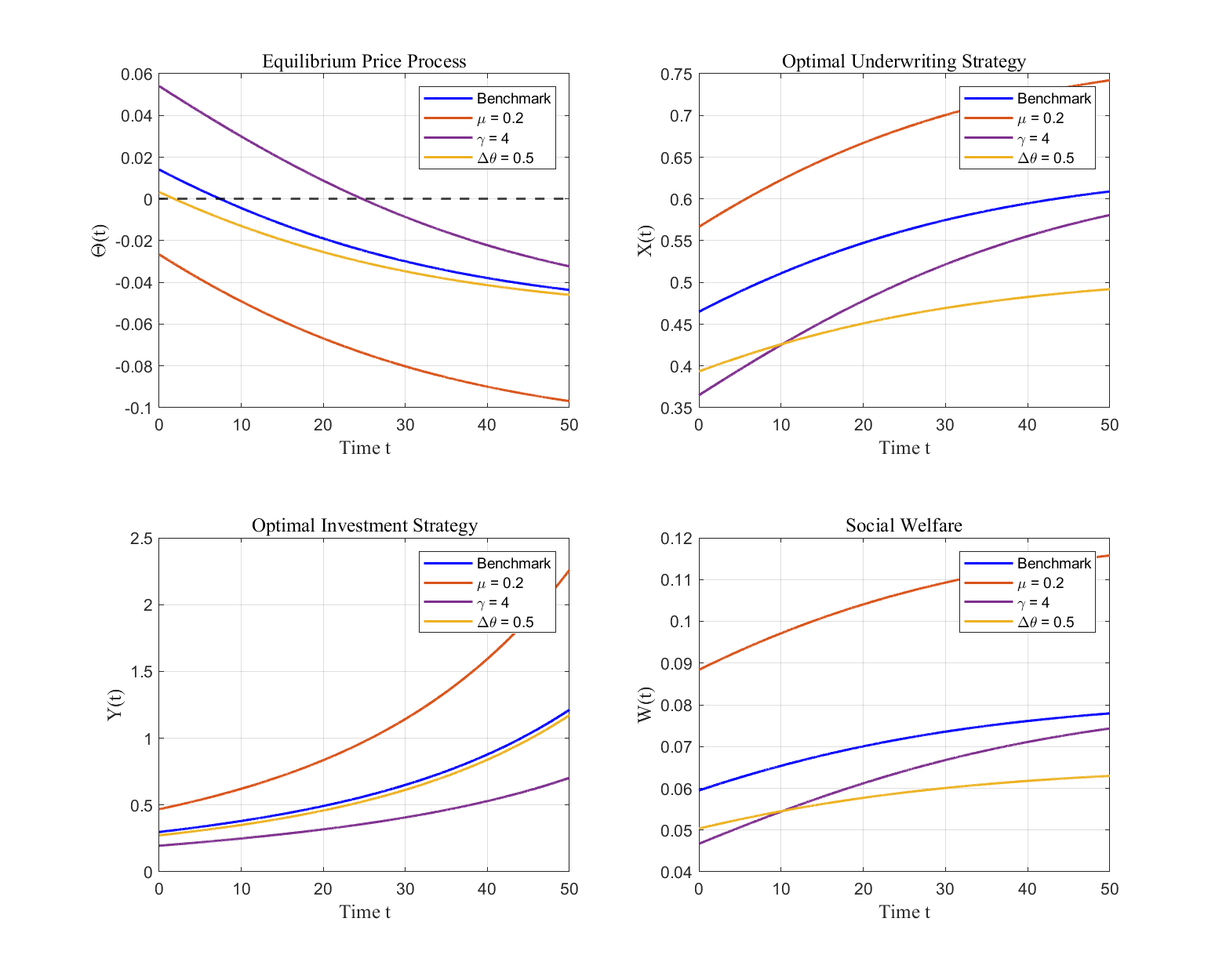}
    \caption{Equilibrium with Negative Correlation}
    \label{Equilibrium with Negative Correlation}
\end{figure} 

We first set $\rho = -0.7$, indicating a negative correlation between insurance and financial markets. Figure \ref{Equilibrium with Negative Correlation} shows that as time progresses, the equilibrium insurance price decreases and eventually becomes negative, consistent with Proposition \ref{Proposition Negative Loading}. The investment amount increases convexly, while the underwriting amount and social welfare increase concavely. Increasing $\mu$ from $0.12$ to $0.2$ raises the Sharpe ratio, which leads to a lower insurance price, higher underwriting, increased investment, and improved welfare, consistent with Proposition \ref{Proposition Comparative Static} and Theorem \ref{Theorem Social Welfare}. When changing $\gamma$ from $2$ to $4$, higher risk aversion leads to a higher price, reduced underwriting and investment, and lower social welfare. Finally, widening the price gap $\Delta \theta$ by lowering $\underline{\theta}$ from $-0.2$ to $-0.3$ leads to a decrease in the insurance price. 

\begin{figure}[htb]
    \centering
    \includegraphics[width=1.0\linewidth]{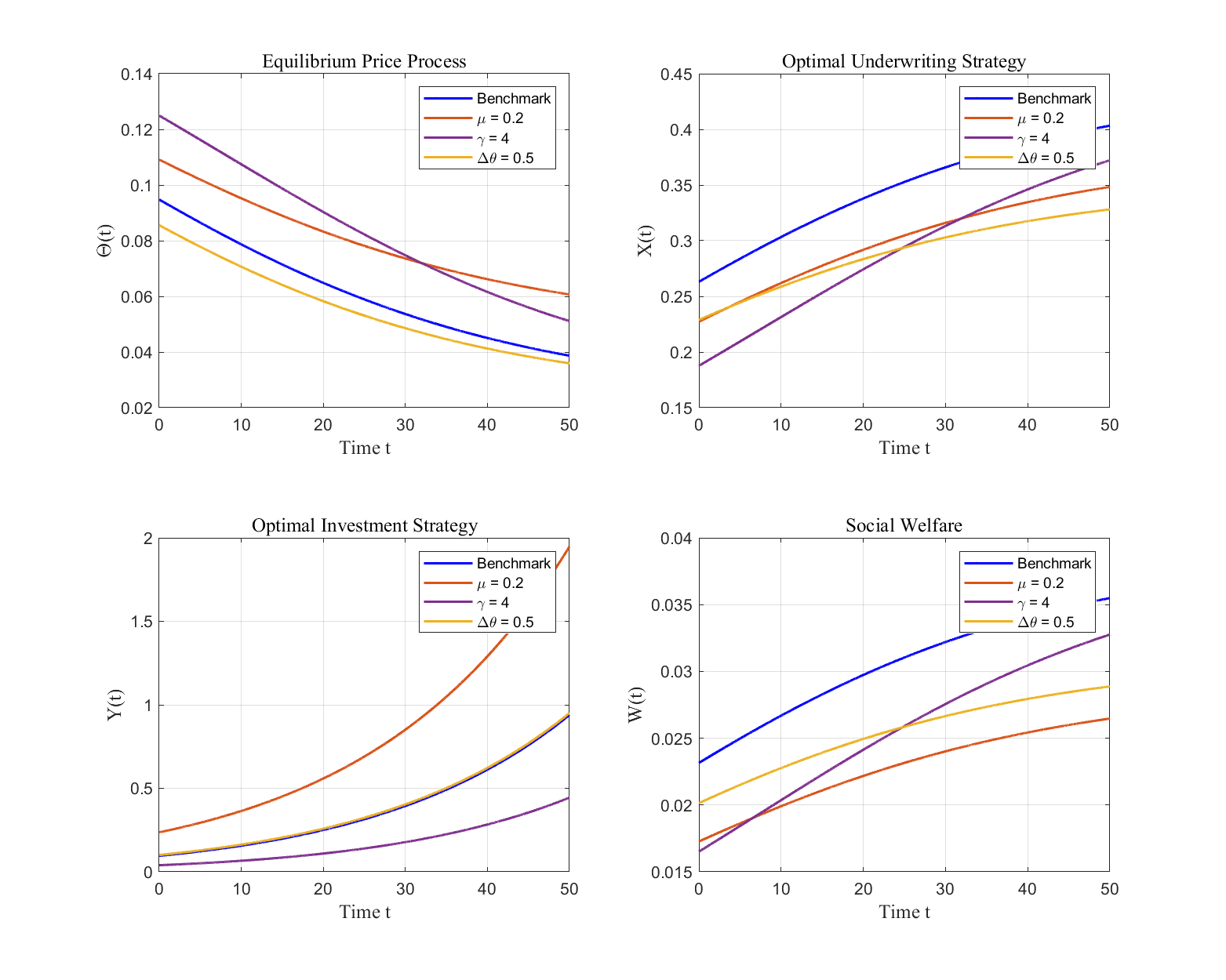}
    \caption{Equilibrium with Positive Correlation}
    \label{Equilibrium with Positive Correlation}
\end{figure}

However, in the case where $\rho = 0.3$, Figure \ref{Equilibrium with Positive Correlation} shows different dynamics compared to the negative correlation scenario. While the equilibrium price remains positive throughout, the increase in the Sharpe ratio leads to reversed effects on the equilibrium price, underwriting amount, and social welfare. 

Now we introduce a regulatory investment cost with an intensity bounded by $\overline{\epsilon} = 0.2$. Under the benchmark parameter setting, we have $\overline{\rho}_s = 0.4 + 4 e^{0.04s - 2}$, $\underline{\rho}_s = -1.25 + \sqrt{3.5625 + 20e^{0.04s - 2}}$, for $s \in [0, 50]$. Clearly, $\overline{\rho}_s \geq \overline{\rho}_0 = 0.941$ and $\underline{\rho}_s \geq \underline{\rho}_0 = 1.254 > 1$. Thus, under this setting, only cases (1) and (2) from Theorem \ref{Theorem Regulated Social Welfare} are relevant, depending on the sign of $\rho$. 

\begin{figure}[htb]
    \centering
    \includegraphics[width=1.0\linewidth]{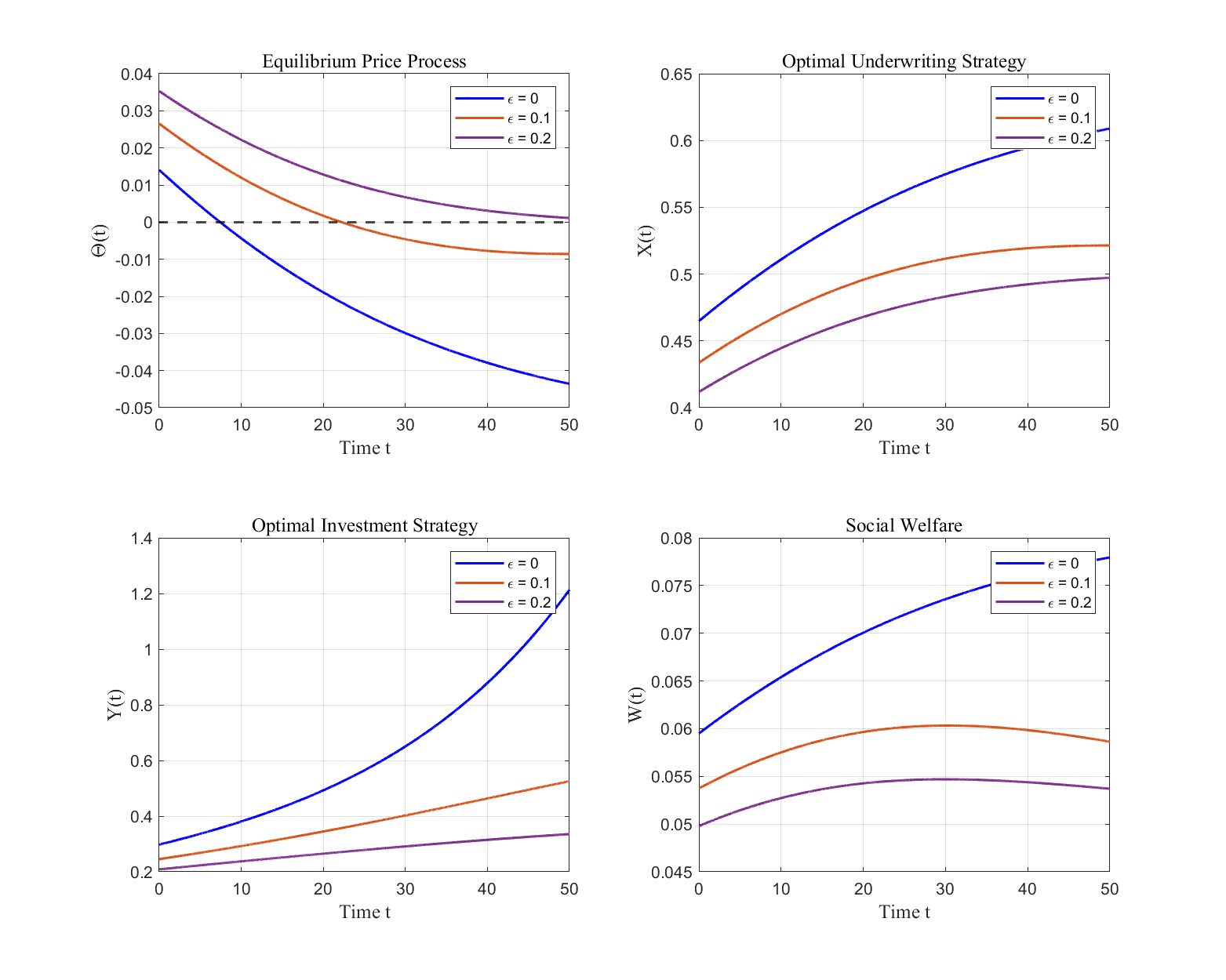}
    \caption{Equilibrium with Regulatory Cost and Negative Correlation}
    \label{Regulatory Equilibrium with Negative Correlation}
\end{figure}

Assume $\epsilon_s = \epsilon$ for any $s \in [0, 50]$. When $\rho = -0.7$, Figure \ref{Regulatory Equilibrium with Negative Correlation} illustrates that an increase in regulatory cost raises the equilibrium insurance price, reduces the underwriting and investment quantities, and leads to lower social welfare. These patterns are consistent with the results from Proposition \ref{Proposition Regulatory Comparative Static} and Theorem \ref{Theorem Regulated Social Welfare}. From the perspective of maximizing social welfare, regulators should avoid imposing any cost.

\begin{figure}[htb]
    \centering
    \includegraphics[width=1.0\linewidth]{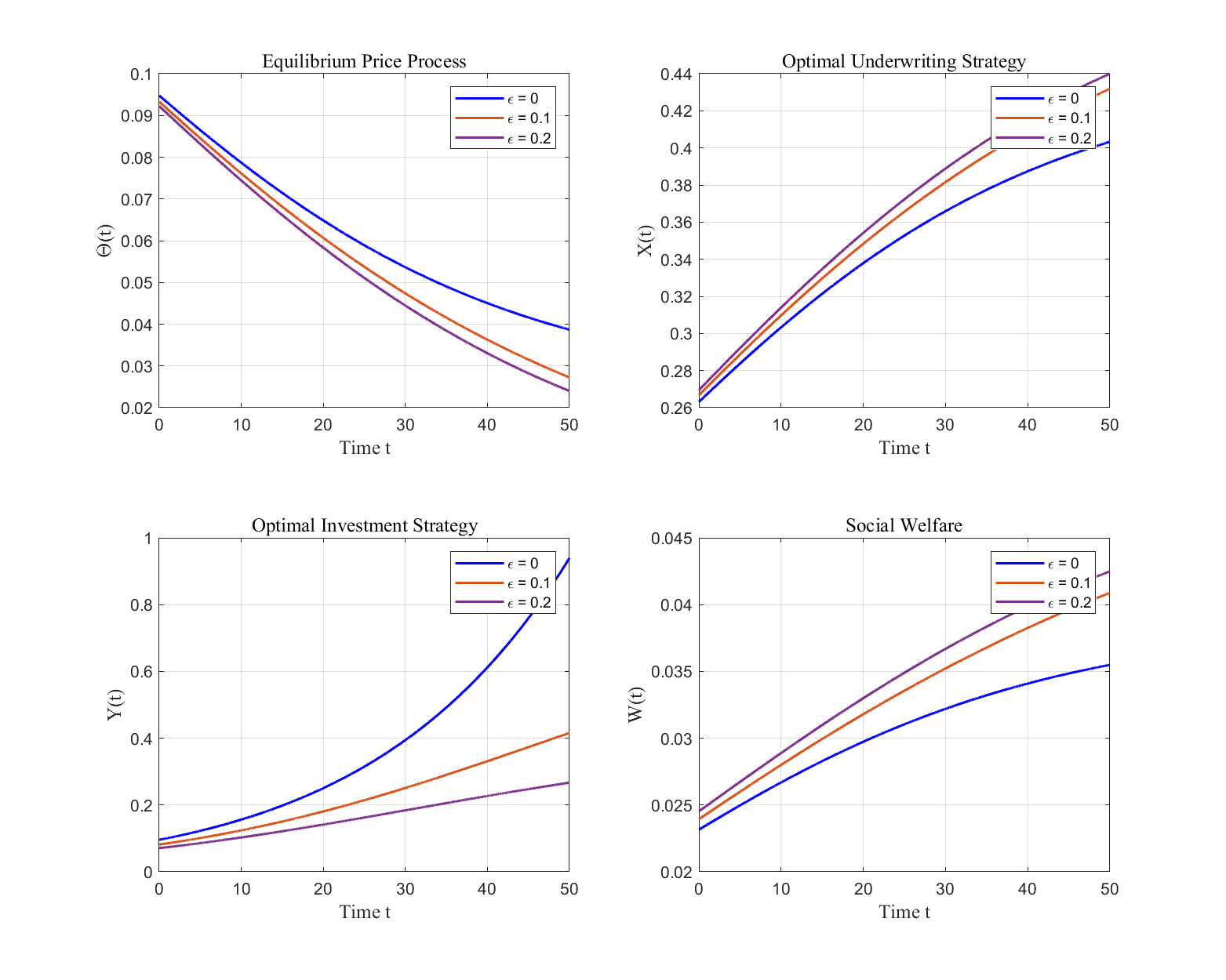}
    \caption{Equilibrium with Regulatory Cost and Positive Correlation}
    \label{Regulatory Equilibrium with Positive Correlation}
\end{figure}

However, when $\rho = 0.3 < \overline{\rho}_s$, as shown in Figure \ref{Regulatory Equilibrium with Positive Correlation}, the increase in regulatory cost leads to reversed effects on the equilibrium price, underwriting amount, and social welfare. In this case, imposing the maximum allowable regulatory cost proves to be the optimal policy, as it stabilizes the market by encouraging more underwriting activity and boosting overall welfare. In a short summary, the regulatory investment cost has asymmetric effects depending on the sign and magnitude of $\rho$. We stress the importance of tailoring regulatory interventions to the specific market conditions and correlations at play. 

\section{Conclusion} \label{Section Conclusion}

In this paper, we develop a dynamic model to study the equilibrium of insurance market, emphasizing the interaction between insurers' underwriting and investment strategies. Depending on the parameter conditions, three distinct equilibrium outcomes can emerge: \textit{positive insurance market}, \textit{zero insurance market}, and \textit{market failure}. Specifically, insurers may rationally accept underwriting losses while relying on investment returns, particularly when there is a negative correlation between insurance gains and financial returns. This provides a theoretical explanation for the observed behavior of Chinese P\&C insurers. 

We also explore the impact of regulatory frictions on market equilibrium and social welfare. While regulatory costs imposed on investments can enhance market efficiency in some cases, a zero-cost approach may be optimal in others. Understanding the magnitude of the correlation between insurance and financial markets is crucial for insurers to formulate optimal strategies and for regulators to design effective policies. While accurately estimating this correlation is challenging in practice \citep{cheng2024robust}, we stress the importance of exploring equilibrium outcomes in the presence of correlation ambiguity in future studies.

\bibliographystyle{apalike}
\bibliography{aguiar}
\end{document}